\documentclass[journal]{IEEEtran}
\usepackage{amsmath,amsfonts,amsthm,amssymb}
\usepackage{algorithmic}
\usepackage{algorithm}
\usepackage{array}
\usepackage{textcomp}
\usepackage{stfloats}
\usepackage{url}
\usepackage{hyperref}
\hypersetup{colorlinks=false}
\usepackage{xcolor,soul,framed} 
\usepackage{verbatim}
\usepackage{graphicx}
\usepackage{makecell}
\usepackage{cite}
\usepackage{soul}
\usepackage{color}
\usepackage{epstopdf}
\usepackage{subfigure}
\usepackage{booktabs}
\usepackage{gensymb}
\usepackage{epstopdf}

\begin{document}
\newtheorem{Remark}{Remark}
\newtheorem{remark}{Remark}
\renewcommand{\algorithmicrequire}{\textbf{Input:}} 
\renewcommand{\algorithmicensure}{\textbf{Output:}}

\title{Multi-RIS-aided  Wireless Communications in Real-world: Prototyping and Field Trials}

\author{Rujing Xiong, Jianan Zhang, Xuehui Dong, Zhengyu Wang, Junshuo Liu,~Wei Yang,~\IEEEmembership{Student Member,~IEEE,}
Tiebin Mi,~\IEEEmembership{ Member,~IEEE,}
~Wenbo Huang,~\IEEEmembership{Student Member,~IEEE,}
Robert Caiming Qiu,~\IEEEmembership{Fellow,~IEEE,}
\thanks{R.~Xiong et al. are with the School of Electronic Information and Communications, Huazhong University of Science and Technology, Wuhan 430074, China(e-mail: rujing@hust.edu.cn).}
\thanks{NFSC, NO.12141107, supports this work.}
\thanks{Manuscript received xxx, 2023.}}



\maketitle

\begin{abstract}
The performance of multiple reconfigurable intelligent surfaces (RISs) receives limited attention in previous studies. This article fills this research gap by investigating the capabilities of multiple RISs in real-world networks. We propose a simplified yet highly scalable sandwich architecture for implementing one-bit unit cells, with the flexibility to accommodate multi-bit unit cells. To effectively control multiple RISs, we present a cost-effective remote-controlling scheme and develop a cloud-based RIS management system. Through a series of four field trials, we demonstrate the effectiveness of multi-hop routing schemes in establishing reliable links. Our experiments reveal significant improvements in signal strength and data transmission in multi-RIS-aided Wi-Fi and commercial 5G networks. Furthermore, we investigate the power scaling law of RIS-aided beamforming and provide insights into the roles of the later nodes in multi-hop relay chains.

\end{abstract}


\section{Introduction}
Reconfigurable intelligent surfaces (RISs) have gained considerable attention due to their ability to intelligently manipulate wireless propagation environments. Unlike conventional relays, RISs are passive in nature and do not require complex RF circuits. A RIS consists of regularly arranged electromagnetic (EM) unit cells, typically sub-wavelength microstrip patches printed on a dielectric substrate~\cite {della2014digital}. These cells are programmable and provide the ability to control the reflection properties of EM waves. This emerging technology holds great potential for next-generation communications and finds applications in various domains~\cite{makarfi2020reconfigurable, kisseleff2020reconfigurable}.

In terms of the architecture of unit cells, RIS can be categorized into two classes: stub-based and slot-based. Both of these architectures feature a top primary metallic patch that is responsible for generating specific resonance characteristics. Regarding the tunability of unit cells, RIS can be classified into positive intrinsic negative (PIN) diode-based and varactor diode-based. 

To assess the practical effectiveness of RISs in real-world scenarios, various experiments have been conducted~\cite{gros2021reconfigurable,tang2020wireless,pei2021ris,dai2020reconfigurable,rains2021high,trichopoulos2022design,sang2022coverage,araghi2022reconfigurable}. The majority of these experiments primarily focus on evaluating the performance of an individual RIS. These measurements are conducted in anechoic chambers~\cite{gros2021reconfigurable,tang2020wireless}~and practical environments~\cite{pei2021ris,dai2020reconfigurable,rains2021high,trichopoulos2022design,sang2022coverage,araghi2022reconfigurable} to investigate fundamental characteristics.
Laboratory equipment such as universal software radio peripherals (USRP), spectrum analyzers, and vector network analyzers are typically employed to facilitate data collection and ensure accurate measurements.

To the best of our knowledge, there are limited field trials evaluating the performance of multiple RISs in real-world commercial networks. Only a few studies explore the characters and behaviors of multiple RISs. In this article, we assess the performance of two-hop RISs in Wi-Fi and commercial 5G networks. Additionally, we present empirical results obtained from multi-hop RISs experiments conducted in an anechoic chamber. Our findings demonstrate that a multi-hop routing scheme utilizing RISs can establish reliable communication links, even in challenging environments such as long pathways and tunnels.

The contributions of this work are summarized as follows:
\begin{itemize}

  \item \textbf{Simplified architectures}. We introduce a simplified yet highly scalable sandwich architecture for implementing one-bit unit cells. This architecture can easily accommodate multi-bit unit cells. The designed RIS, operating at 5.8 GHz and consisting of $10 \times 16$ such one-bit unit cells, has a maximum power consumption of 0.2 W.

  \item \textbf{Wireless controlling}. To enable remote control of multiple RISs, we utilize a dedicated microcontroller unit (MCU) that integrates Wi-Fi capabilities for networking, this scheme enables the RISs to be controlled either through the Internet or a local area network (LAN). Additionally, we develop a cloud-based management system that leverages the computational power and storage capacity of the servers, resulting in improved scalability.

  \item \textbf{Multi-hop field trials}. We conduct the world's first two-hop field trials in Wi-Fi and 5G networks, highlighting the effectiveness of multiple RISs in coverage enhancement and data transmission. Furthermore, we conduct the world's first multi-hop RIS tests in an anechoic chamber to investigate their fundamental capabilities. We observe limited power gains of the later RIS nodes, providing insights into the practical limitations through multi-hop setups.

\end{itemize}

The remainder of this paper is organized as follows. Section~\ref{Section2} discusses the designs of the RIS, including the stub-based unit cell and the wireless implementation for real-time control of multiple RISs. In Section~\ref{Section3}, we describe the setup of four different multi-RIS-aided communication systems, along with the presentation of the corresponding testing results. Finally, Section~\ref{Section4} concludes the article.

\section{Prototype Designs}\label{Section2}

\subsection{A Simplified Architecture for Unit Cells}
The implementation of unit cells involves the integration of microstrip antennas and tunable components. We first present a simplified yet highly scalable sandwich architecture for the implementation of unit cells. This architecture comprises a primary metallic patch, which serves as the main element to produce certain resonance features, along with switching stubs consisting of microstrip lines and PIN diodes. By controlling the operation state of the PIN diodes, we can achieve distinct reflection phase responses.

\begin{figure*}
  \centering
  \includegraphics[width=0.9\linewidth]{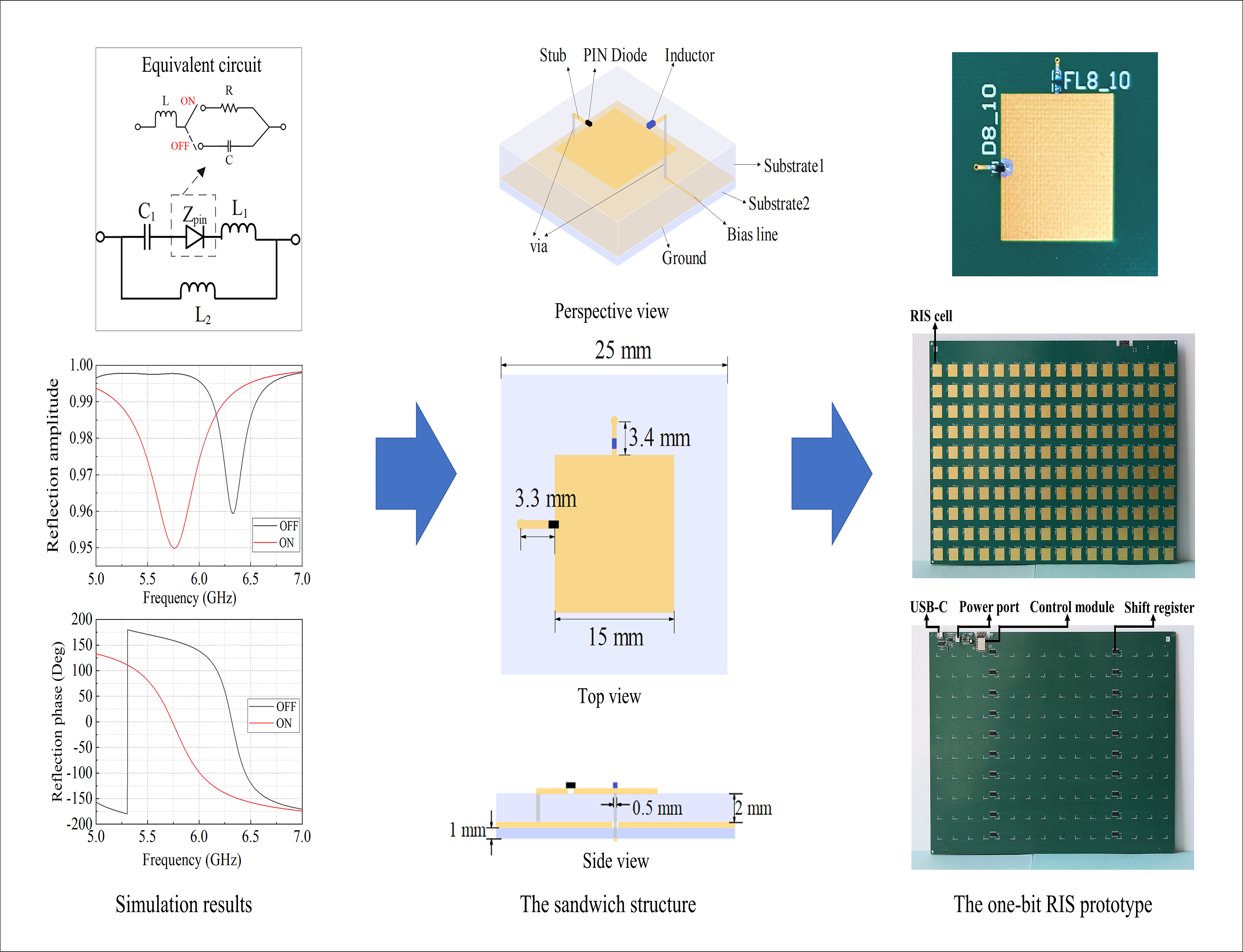}
  \caption{The proposed one-bit RIS operating at center frequency 5.8 GHz.}\label{F1}
\end{figure*}

In the example demonstrated in Fig.~\ref{F1}, we provide a comprehensive overview of the architecture and parameters of a one-bit unit cell operating at 5.8 GHz. The structure consists of three copper layers, with two substrate layers positioned in between. The top sandwich architecture, which consists of the upper two copper layers and substrate, is utilized for implementing the radiated cells. The bottom layer is dedicated to the digital control circuit.

Within the radiated unit cell, the main metallic patch is connected to the bias line. To ensure safety, a choke inductor (LQW15AN19NG00D) is incorporated to isolate the direct current from bias circuits. The adjacent edge of the patch is connected to the ground through a stub. The stub consists of a specially designed microstrip line and a PIN diode (SMP1345-079LF), which enables the phase configuration functionality. The upper substrate is made of F4B with a relative permittivity of 2.65 and a loss tangent of 0.001. The lower substrate is made of the commonly used FR4 material.

The reflection coefficient of the unit cell can be determined by employing equivalent circuit models, which utilize the principles of transmission line theory to transform the complex EM problem into a simpler circuit problem. In the case of the PIN diode, its behavior can be modeled using a lumped-parameter circuit, as illustrated in Fig.~\ref{F1}. When the diode is biased, it is represented by a 0.7 nH inductor in series with a $2 ~\Omega$ resistor. If the diode is switched OFF, it is modeled by a 1.8 pF capacitor in series with a 0.7 nH inductor.

To accurately evaluate the response of the unit cell, we employ CST Studio Suite, a high-performance package for analyzing EM
components. From the simulation results shown in Fig.~\ref{F1}, we observe that the unit cell architecture yields a reflection coefficient exceeding 0.95, indicating that a significant portion of the incident wave is reflected. In addition, the unit cell exhibits two distinct states when configured as ON and OFF, with phase values of $-25^{\circ}$ and $156^{\circ}$, respectively.

It is worth noting that the proposed architecture can be easily extended to accommodate two-bit unit cells. This can be achieved by increasing one more stub. Leveraging this approach, we design a RIS board consisting of $10\times 20$ two-bit unit cells operating at 3.4 GHz. To promote knowledge sharing, the designs of one-bit unit cells operating at 5.8 GHz and RIS prototype consisting of $10 \times 16$ such unit cells are open-sourced at https://github.com/RujingXiong/OpenRIS.git under the name OpenRIS.

\subsection{Wireless Implementation for Controlling}
To effectively control multiple RISs deployed over a wide area, we develop a convenient remote-controlling scheme using cost-effective MCUs and shift registers. Specifically, the RIS board utilizes the ESP8266-12F as the primary control unit and the 74LS595D as the shift register. To facilitate cascading of multiple RIS, the board is designed to operate in two modes: master mode and slave mode. In the slave mode, the MCU enters a low-power sleeping status.

Regarding power consumption, an individual RIS board, which consists of $10 \times 16$ one-bit unit cells operating at 5.8 GHz, exhibits a low power consumption of 0.2 W when operating in master mode. If the RIS board switches to slave mode, the power consumption reduces significantly to only 0.03 W. 

To enable seamless control of multiple RISs, we leverage the integrated Wi-Fi capability of the ESP8266-12F. This built-in feature empowers us to remotely configure multiple RISs operating at different frequencies. We develop a cloud-based RIS management system that utilizes the message queuing telemetry transport (MQTT) protocol. The system incorporates dedicated servers, offering an efficient and scalable solution for managing and controlling a large number of RIS boards across a wide area.

\subsection{Beam Reshaping}
One notable capability of RIS is to redistribute incident power to produce a desired power radiation pattern. Nevertheless, achieving complex beam reshaping functionality is challenging due to the non-convex constraints, particularly the unit modulus constraints on the configurations. To tackle these, several approaches are employed, including semidefinite relaxation~\cite{wu2019intelligent}, alternating direction method of multipliers~\cite{ning2020beamforming}, and manifold optimization (MO)~\cite{xiong2023ris}.

Among the approaches mentioned, MO is particularly recommended. By leveraging the geometry of a manifold, a class of constrained optimization problems can be transformed into unconstrained optimization problems on the manifold. In the case of the unit modulus constraints of configurations, MO directly restricts the solution on a smooth Riemannian manifold embedded in $\mathbb{C^N}$. We employ this method for beam reshaping in the far field \cite{xiong2023ris}.

We now consider the near-field scenario, where the source is located in the near field of the RIS board but still in the far field for each individual unit cell. In this situation, the EM waves propagate as spherical waves. The incident wave illuminating each unit cell introduces a phase delay that is directly proportional to the distance. To achieve a collimated beam, it is necessary for the phase configuration to compensate for this spatial phase difference.

The required phase configurations are given as~\cite{stutzman2012antenna}
\begin{equation}\label{PH}
\begin {aligned}
\omega_i
&=\omega_i^{\rm{Arr}}+\omega_i^{\rm{Dep}}\\
&=-k_0(d_i-\sin \theta_0(x_i \cos \varphi_0+y_i \sin \varphi_0)) .
\end {aligned}
\end{equation}
In this equation, $\omega_i^{\rm{Arr}}$ represents the compensation of the incident phase delay,  $d_i$ is the distance from the source to the unit cell located at $(x_i,y_i)$, and $k_0$ is the wavenumber. The term $\omega_i^{\rm{Dep}}$ represents a progressive phase used to steer the main beam towards the desired direction ($\theta_0, \phi_0$).

\section{Experimental Evaluations}\label{Section3}

The relationship between the number of RIS cells and their performance gains, as well as the deployment strategies of multiple RISs, remains a prominent research topic. We conduct a series of four experiments to explore the advantages and limitations of various RIS-aided wireless communications. These experiments are divided into two approaches: two experiments conducted in anechoic chambers and two experiments conducted in practical environments. In the anechoic chamber experiments, we investigate fundamental characteristics such as scaling laws and limitations of multi-hop routing. Additionally, we evaluate the performance of two RISs in real networks by conducting tests in RIS-aided Wi-Fi 6 and commercial 5G networks under different scenarios.

\subsection{The Bigger the Better}

The scaling law is crucial in designing RIS-aided systems. Although several theoretical works explore the problem, there is still a need for practical experiments to validate and test the scaling law in real-world scenarios. The first experiment in our study is to address this gap by conducting tests using an actual RIS-aided system. We investigate the relationship between the received power and the number of RIS cells. To achieve this, we conduct tests utilizing a single RIS, a dual-RIS, and a quad-RIS for signal enhancement in both the near and far fields.

In our experimental setup, the single RIS board is composed of $10 \times 16$ unit cells. The dual-RIS consists of $10 \times 32$ unit cells, and the quad-RIS has dimensions of $20 \times 32$. We employ a full duplex software-defined-radio USRP 2954R along with a pair of horn antennas in the anechoic chamber. The carrier frequency for the tests is set at 5.8 GHz, and the transmitted signal has a bandwidth of 100 KHz. The received signal power is calculated by averaging 8192 data points.

In the near-field test, optimizing the configurations of unit cells leads to significant improvements in received power. When a single RIS is configured, the received power increases from -65 dBm to -43 dBm. Activating the dual-RIS enhances the received power to -33 dBm. Deploying the quad-RIS leads to an increase in received power to -28 dBm. These findings provide clear evidence of the substantial improvement in received signal power achieved through the utilization of RISs.

We conduct the same tests in the far-field region. Similar to the near-field results, we observe significant improvements in received signal power by optimizing the configurations of a single RIS, dual-RIS, and quad-RIS. After optimal configuration, the measured received power levels are as follows: -39.48 dBm for the single RIS, -36.03 dBm for the dual-RIS, and -33.41 dBm for the quad-RIS. These results provide valuable insights into the relationship between received power and the number of unit cells. Specifically, in the far-field region, we observe that doubling the number of unit cells leads to an approximate gain of 3 dB in received power.  This finding suggests a received power scaling law with respect to the number of RIS cells.

In summary, our practical experiments reveal a positive relationship between the received power and the size of the RIS in both near-field and far-field tests. Increasing the number of unit cells leads to higher power gains, highlighting the importance of RIS dimensions. However, the power scaling law is more complex in the near-field, due to the characteristics of the Fresnel zone, making it challenging to establish a clear linear relationship.

\begin{figure*}
\centering
  \includegraphics[width=1\linewidth]{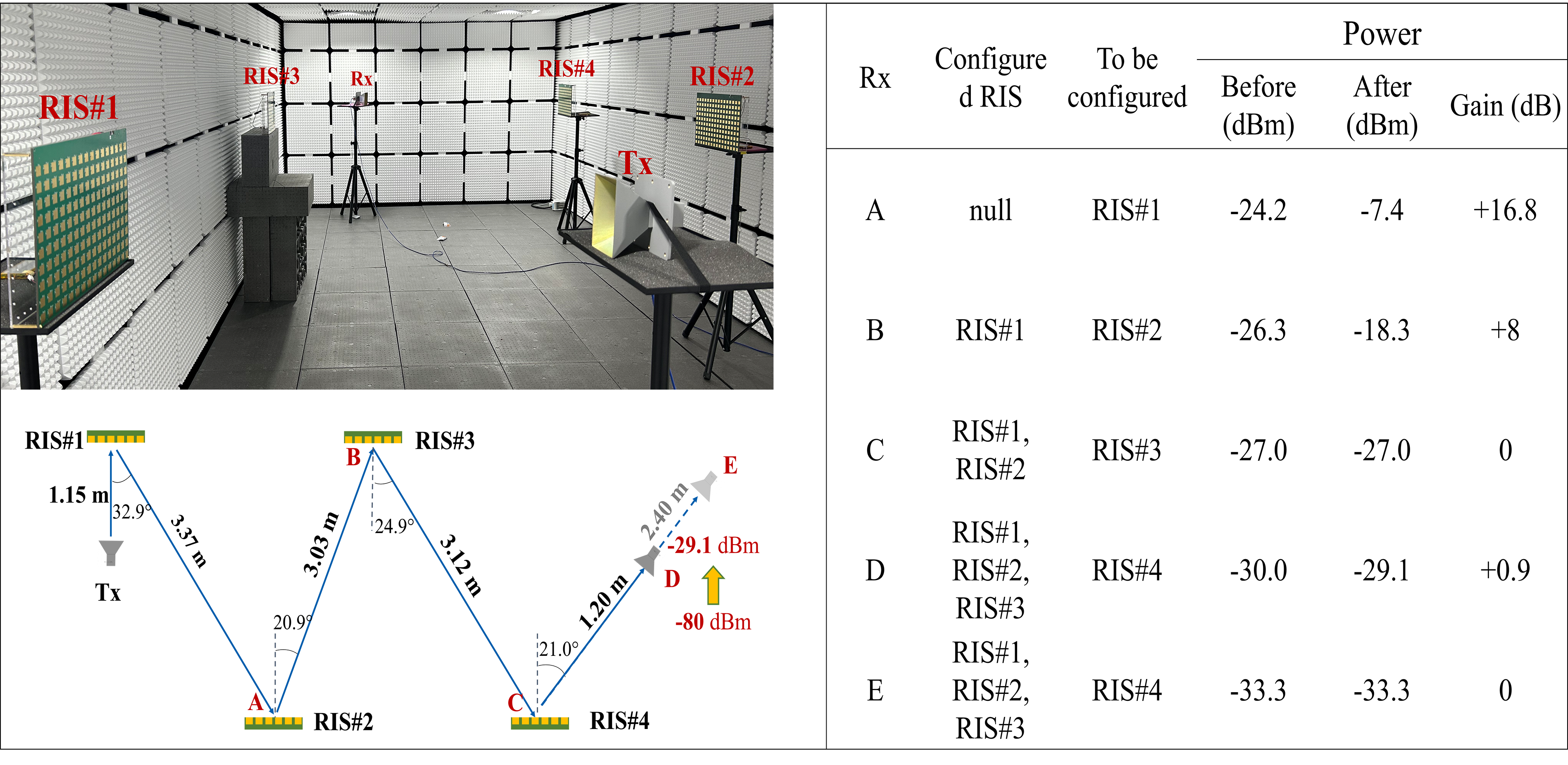}
\caption{Experiments of multi-hop RISs-aided communication.}
\label{F2}
\end{figure*}

\subsection{Multi-hop Routing}

The effectiveness of multi-RIS-aided communication, particularly involving more than two hops, has not been thoroughly explored in practical experiments. To address this, we conduct experiments in the anechoic chamber following the setup illustrated in Fig.~\ref{F2}. Multiple RISs, with a maximum of four, are utilized to relay signals from the source to the receiver in a sequential manner. Each RIS comprises $10 \times 16$ one-bit unit cells and is powered by a Li-ion battery.

In the first experiment, we investigate the power gain achieved by different RIS routing setups, including one-hop, two-hop, three-hop, and four-hop routes. The source emits EM waves towards RIS\#1 at an angle of $0^{\circ}$. Beam reshaping techniques are employed at each RIS to focus the incident EM waves towards the subsequent RIS location. To ensure fair comparisons, the horn antenna of the receiver is directed towards the last RIS in each test.

Without the presence of RISs, the power level at the receiver is measured to be approximately -80 dBm. When the RISs are strategically positioned and optimally configured, we observe significant improvements in the received power at the receiver. The power levels measured are -13.9 dBm, -23.3 dBm, -26.4 dBm, and -29.1 dBm for the one-hop, two-hop, three-hop, and four-hop RISs routes, respectively. These results indicate that the RISs can establish reliable links between the transmitter and the receiver. 

An interesting finding is that the power gain achieved by multi-hop RISs decreases as the number of RISs in the relay chain increases. This trend can be attributed to the path losses introduced by multiple path segments due to the presence of multiple RISs. While the RISs contribute gains, these gains may not fully compensate for the cumulative path losses along the relay chain. Therefore, it is preferable to use fewer RISs, if possible, to minimize the impact of additional path losses.

In the second study, we conduct additional investigations on the multi-hop RISs routing along a designated path. Unlike the first experiment, where the receiver's position is fixed and oriented towards the last RIS, we change the receiver's location. The receiver is currently positioned at points A, B, C, and D (or E) for the one-hop, two-hop, three-hop, and four-hop RISs routes, respectively. 

The results demonstrated in Fig.~\ref{F2} indicate the distinct roles played by each RIS in enhancing the signal. Specifically, RIS\#1 (positioned at the beginning of the chain) provides a substantial power gain of 16.8 dB, making it the most influential in enhancing the received signal. RIS\#2 also contributes a gain of 8 dB, although not as significant as the first. Surprisingly, even with optimal configurations, RIS\#3 and RIS\#4 do not exhibit substantial power gains. However, these later RIS nodes play a role in extending the propagation path.

To gain further insights into the latter nodes, an additional test is conducted to explore the role of RIS\#3. We replace the single RIS with a quad-RIS comprising $20 \times 32$ unit cells. The objective is to assess whether a larger RIS can enhance the overall performance. However, the results show that the power gain achieved by the quad-RIS is only 0.9 dB, which is significantly lower than the theoretical value of 6 dB. The main reason behind this observation is the limited reflective power from RIS\#2, resulting in a weak incident wave impinging on the next  RIS\#3. Despite configuring RIS\#3 optimally, the reflective power is still weak.

In summary, the experiments yield two interesting findings:
\begin{itemize}
\item Multi-hop RISs effectively enhance wireless communication in challenging environments, such as tunnels, resulting in significant improvements in received signal power.

\item The power gain achieved by multi-hop RISs varies depending on the number of RISs in the relay chain. While multiple RISs contribute gains to enhance signals, the power gains provided by the latter RISs may not fully compensate for the introduced path losses.
\end{itemize}

\begin{figure*}[!htbp]
  \centering
  \includegraphics[width=1\linewidth]{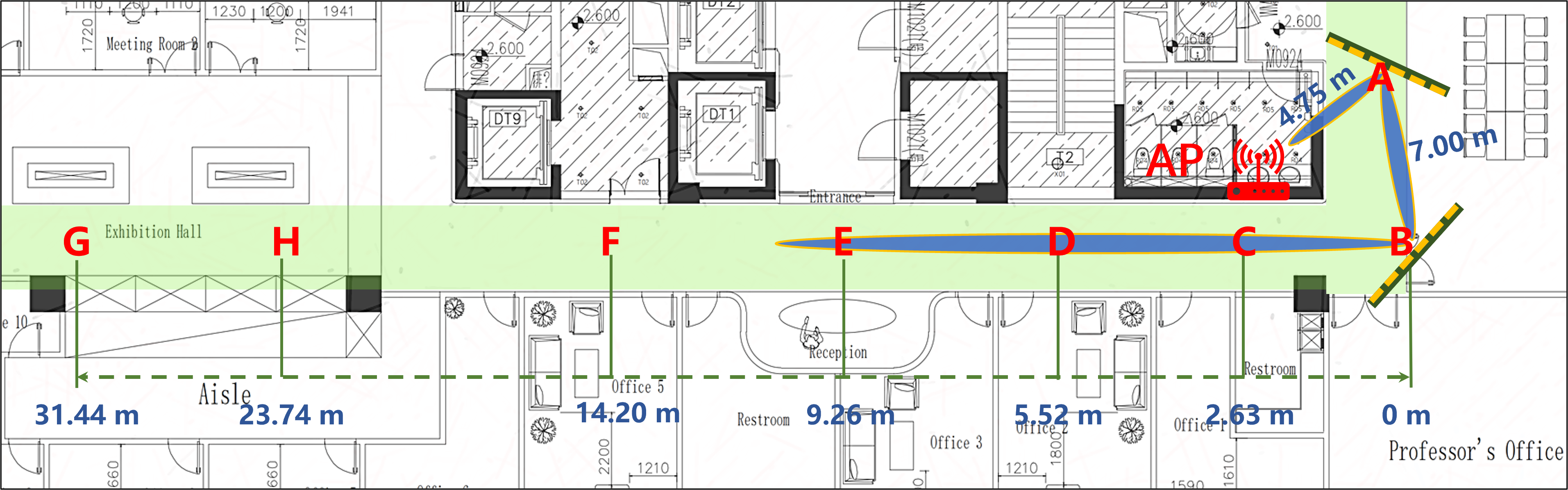}
  \caption{Experiment setup of Multi-RIS-aided Wi-Fi communication in an office building.}\label{F3}
\end{figure*}

\subsection{RIS-aided Wi-Fi Communications}
In order to evaluate the performance of the multiple RISs in the real network, we conduct tests in a RIS-aided Wi-Fi 6 network. The experimental setup consists of an access point (HONOR XD20 @5.8 GHz) serving as the source and a station (Lenovo ThinkPad T14) acting as the receiver. One or two RISs are deployed in the link to assess their impact on the network performance. In addition to the signal strength, we also consider other metrics such as data rate, jitter, and packet loss as evaluation criteria. These metrics provide a comprehensive assessment of the network's performance, specifically in relation to user experience.

To evaluate the capability of signal enhancement, we perform an experiment using a single RIS to relay Wi-Fi signals into an anechoic chamber of size $5 \times 7 \times 6 \ \rm{m^3}$. The access point is positioned outside the anechoic chamber, while a RIS operating at 5.8 GHz is placed at the entrance. To measure and analyze the signal strength, we utilize the Wi-Fi analyzer tool called NetSpot. On average, there is an improvement of over 10 dB for signal strength in the anechoic chamber. This enhancement is directly attributed to the RIS's capability to reshape the incident signal, enhancing coverage and mitigating the blind spot

Furthermore, we construct a two-hop RISs setup to enhance Wi-Fi 6 coverage along the L-shaped pathway in an office building.  In addition to utilizing NetSpot to assess signal strength, we also use iPerf3 to evaluate the performance in terms of data rate, jitter, and packet loss. iPerf3 is a standardized performance measurement tool because it provides accurate measurements that closely relate to user experience.

Fig.~\ref{F3} illustrates the setup of the two-hop RIS-aided Wi-Fi network. The access point acts as the transmitter, located in the restroom and directed towards the first RIS in the short hallway. The second RIS, positioned at the corner, acts as a relay to enhance the Wi-Fi signal along the long hallway. Both RISs are positioned at a height of 1 m. We optimize the phase configurations using the RGD algorithm\cite{xiong2023ris}.

The performance indicators at eight different points (A to H) are recorded using a laptop and presented in Table \ref{Tab1}. Notably, as the distance increases, there is a significant improvement in data rate, jitter, and packet loss. The far end points (H and G) in the long hallway exhibit the most significant improvement. For example, at point H, the data rate increases from 17.27 Mbps to 34.80 Mbps, the jitter decreases from 52.56 ms to 9.91 ms, and the packet loss decreases from 69.20\% to 7.51\%. In contrast, the near end points (A and B) show only a modest improvement. This can be attributed to the fact that the performance at these points is already good enough even without the presence of the RISs.

\begin{table*}[!t]
\caption{User experience performance indicators at different points}
\label{Tab1}
\centering
\setlength{\tabcolsep}{1.8mm}
\begin{tabular}{cccccccccc}
\toprule
Locations & \multicolumn{2}{c}{Rx Signal Strength (dBm)} & \multicolumn{2}{c}{Data Rate/Mbps}& \multicolumn{2}{c}{Jitter/ms}  & \multicolumn{2}{c}{Loss/\%} \\
\midrule
Identifier&without RIS &with RIS &  without RIS & with RIS & without RIS & with RIS & without RIS & with RIS\\

A & -55 & -53 & 40.00 & 40.00  & 4.89 &3.22  &0.00  &  0.00 \\

B & -56 & -54 & 40.00 & 40.00 &  7.00 &3.45  & 0.92 &  0.00 \\

C & -66 & -63 & 39.70 & 40.00& 10.08  &3.78   & 5.70 & 0.03  \\

D & -71 & -66 & 36.66 & 40.00&  9.28 &7.76   & 13.52 & 0.03  \\

E &-76  & -69 & 34.62 & 39.90&  25.65 &7.57 & 41.00 & 2.33   \\

F &-78  & -72 & 30.68 & 39.30&  27.61 &8.74   &48.95  & 6.94  \\

G &-79  & -72 & 27.73 & 36.50&  29.93  &8.82   & 63.50 & 7.20  \\

H &-83  & -73 &17.27 &34.80 &  52.56  &9.91   & 69.20 &7.51   \\
\bottomrule
\end{tabular}
\end{table*}

\subsection{RIS-aided Cellular Communications}

\begin{figure*}[h]
  \centering
  \includegraphics[width=1.0\linewidth]{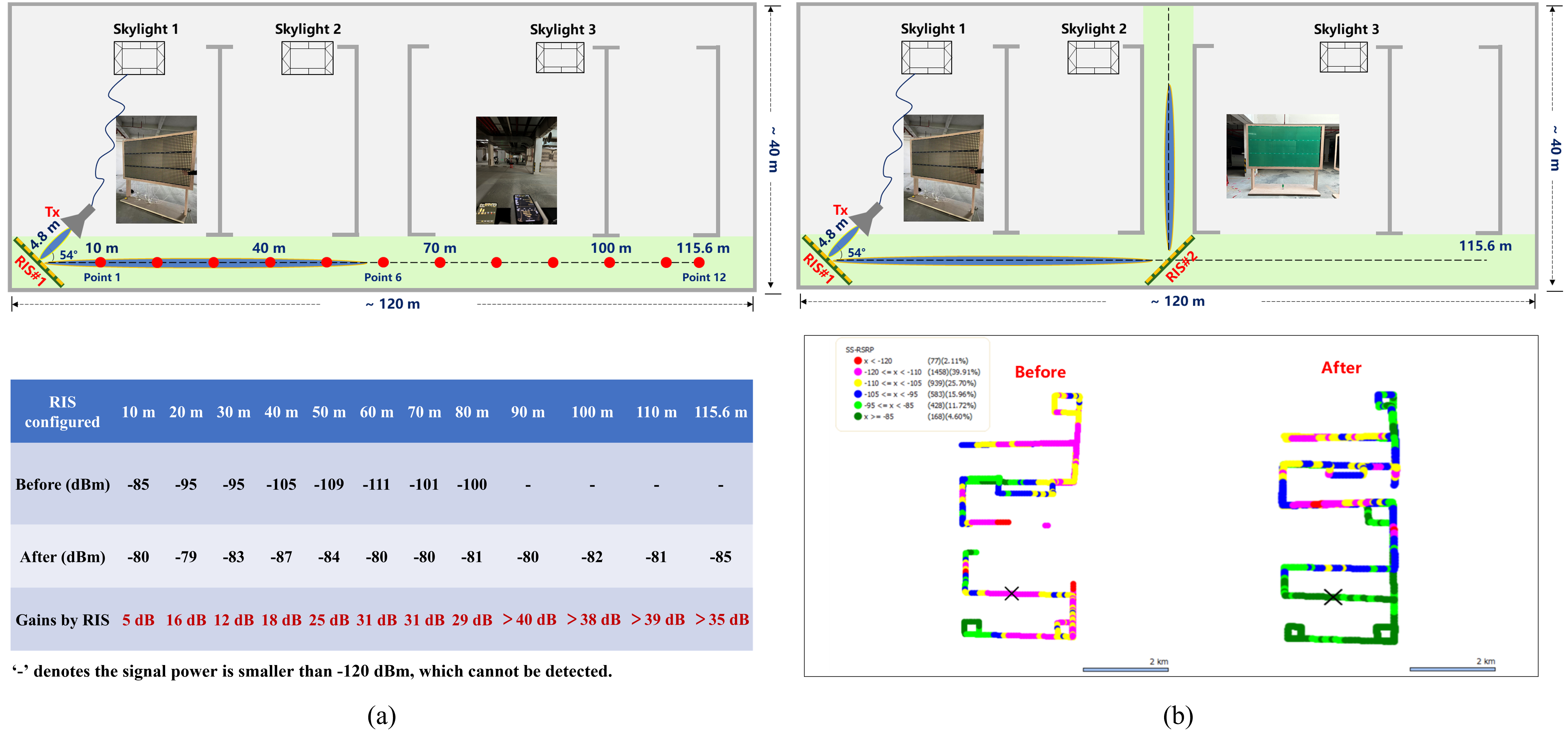}
  \caption{Multi-RIS-aided Cellular Communications. (a) Evaluating signal enhancement along the pathway using a single RIS located at the bottom left corner. (b) Evaluating signal enhancement throughout the entire parking garage using two-hop RISs.}\label{F4}
\end{figure*}

In order to explore the potential of multiple RISs in real-world commercial 5G networks, we perform additional evaluations in an underground parking garage. The parking garage has an approximate area of $125$~$\rm{m} \times40~\rm{m}$ and presents challenges such as signal blockage and path loss, which can have a substantial impact on cellular communication. Our purpose is to evaluate the effectiveness of deploying RISs within the garage to improve signal coverage and enhance overall network performance.

To facilitate the experiment, we utilize a custom-designed repeater that captures signals at the N78 band (3.4-3.5 GHz) of China Telecom. The signal enters the parking garage through the skylight, as illustrated in Fig.~\ref{F4} (a), and the repeater serves as the feed source. It is important to note that even with the repeater activated, the strength of the N78 band in the underground parking remains extremely weak. The average reference signal receiving power (RSRP) falls below the acceptable threshold (-100 dBm) set by operators. In order to improve the signal quality, we strategically deploy two RISs operating at 3.4 GHz, each consisting of $30 \times 60$ two-bit unit cells. The mobile devices act as the receivers.

We conduct two trials to assess the performance of the RIS-aided network. In the first trial, we deploy a single RIS at the bottom left corner to evaluate the signal enhancement along the pathway. We collect measurement data at 12 points using a smartphone and Cellular-Pro, a free professional network optimization app. In the second trial, an additional RIS is deployed at a T-junction to examine the overall performance improvement across the entire parking garage. To ensure comprehensive and accurate measurements, we utilize smartphones and the Pilot Pioneer software, a commercial tool for network performance benchmarking. These measurements offer valuable insights into subscribers' perceptions of the services provided.

The results of the first trial are illustrated in Fig.~\ref{F4} (a) (bottom). After optimal configurations of RIS, the average signal strength along the 120-meter pathway improves from -100 dBm to -80 dBm, resulting in a gain of around 20 dBm. Furthermore, we observe that the RSRP remains around -80 dBm along the entire length of the entire pathway.

The improvement in signal strength can be attributed to two factors. Firstly, at the receiving stage, increasing the number of unit cells boosts the received power, allowing the passive RIS to convert and redirect more power towards the desired direction. Secondly, at the reflecting stage, the large area of the RIS contributes to a higher beamforming gain. That means the larger aperture enables a sharper and stronger main beam. This feature further enhances signal coverage along the pathway.

In the second trial, we evaluate the performance of two-hop RIS-aided cellular communications. To achieve a higher coverage rate, we strategically position the second RIS at a T-junction, as depicted in Fig.~\ref{F4}(b). In contrast to the first trial where only 12 points along the pathway are sampled, our goal is to collect comprehensive signal measurements throughout the entire parking garage. To accomplish this, we employ three smartphones and make use of the Pilot Pioneer software for real-time measurements.

The measurement results from the second trial exhibit remarkable enhancements in signal coverage across the entire parking garage. As shown in Fig.~\ref{F4}(b) (bottom), there is a significant reduction in the region with signal strengths below -110 dBm (represented by the red and pink points) and a significant increase in the region with signal strengths above -95 dBm (represented by the light and dark green points). The average RSRP in the entire garage shows an improvement from -105.84 dBm to -91.87 dBm, resulting in a gain of approximately 14 dB. Additionally, the downlink speed experiences an improvement from 79.63 Mbps to 110.58 Mbps, while the uplink speed increases from 9.99 Mbps to 17.54 Mbps. These findings indicate a significant enhancement in both signal quality and data transmission rates within the two-hop RIS-aided 5G network.

\section{Conclusion}\label{Section4}
This study focuses on the design and performance evaluation of multi-RIS-aided communications. We propose a simplified and scalable architecture for the implementation of unit cells and fabricate RIS boards operating at 5.8 GHz and 3.4 GHz. Through a series of experiments conducted in anechoic chambers and practical environments, we assess the performance of the designed RISs and investigate the potential benefits of multiple RISs.

Our findings indicate that multi-hop RISs effectively enhance wireless communication in challenging environments, resulting in significant improvements in received signal power. Moreover, in certain scenarios, the power gain achieved by optimally configured RISs is directly proportional to the number of unit cells, following a power scaling law. However, an interesting observation is that the power gain can vary considerably along the relay chain of multi-hop RISs, with later RISs experiencing lower power gains compared to earlier nodes. Future research can concentrate on developing collaborative algorithms among multiple RISs.


%

\bibliographystyle{ieeetr} 
\bibliography{Reference}

\begin{thebibliography}{10}

\bibitem{della2014digital}
C.~Della~Giovampaola and N.~Engheta, ``Digital metamaterials,'' {\em Nature
  Materials}, vol.~13, no.~12, pp.~1115--1121, 2014.

\bibitem{makarfi2020reconfigurable}
A.~U. Makarfi {\em et~al.}, ``Reconfigurable intelligent surface enabled {IoT}
  networks in generalized fading channels,'' in {\em ICC 2020-2020 IEEE
  International Conference on Communications (ICC)}, pp.~1--6, IEEE, 2020.

\bibitem{kisseleff2020reconfigurable}
S.~Kisseleff {\em et~al.}, ``Reconfigurable intelligent surfaces for smart
  cities: Research challenges and opportunities,'' {\em IEEE Open Journal of
  the Communications Society}, vol.~1, pp.~1781--1797, 2020.

\bibitem{gros2021reconfigurable}
J.-B. Gros {\em et~al.}, ``A reconfigurable intelligent surface at mmwave based
  on a binary phase tunable metasurface,'' {\em IEEE Open Journal of the
  Communications Society}, vol.~2, pp.~1055--1064, 2021.

\bibitem{tang2020wireless}
W.~Tang {\em et~al.}, ``Wireless communications with reconfigurable intelligent
  surface: Path loss modeling and experimental measurement,'' {\em IEEE
  Transactions on Wireless Communications}, vol.~20, no.~1, pp.~421--439, 2020.

\bibitem{pei2021ris}
X.~Pei {\em et~al.}, ``{RIS}-aided wireless communications: Prototyping,
  adaptive beamforming, and indoor/outdoor field trials,'' {\em IEEE
  Transactions on Communications}, vol.~69, no.~12, pp.~8627--8640, 2021.

\bibitem{dai2020reconfigurable}
L.~Dai {\em et~al.}, ``Reconfigurable intelligent surface-based wireless
  communications: Antenna design, prototyping, and experimental results,'' {\em
  IEEE Access}, vol.~8, pp.~45913--45923, 2020.

\bibitem{rains2021high}
J.~Rains {\em et~al.}, ``High-resolution programmable scattering for wireless
  coverage enhancement: An indoor field trial campaign,'' {\em arXiv preprint
  arXiv:2112.11194}, 2021.

\bibitem{trichopoulos2022design}
G.~C. Trichopoulos {\em et~al.}, ``Design and evaluation of reconfigurable
  intelligent surfaces in real-world environment,'' {\em IEEE Open Journal of
  the Communications Society}, vol.~3, pp.~462--474, 2022.

\bibitem{sang2022coverage}
J.~Sang {\em et~al.}, ``Coverage enhancement by deploying ris in 5g commercial
  mobile networks: Field trials,'' {\em IEEE Wireless Communications}, 2022.

\bibitem{araghi2022reconfigurable}
A.~Araghi {\em et~al.}, ``Reconfigurable intelligent surface ({RIS}) in the
  sub-6 {GHz} band: Design, implementation, and real-world demonstration,''
  {\em IEEE Access}, vol.~10, pp.~2646--2655, 2022.

\bibitem{wu2019intelligent}
Q.~Wu and R.~Zhang, ``Intelligent reflecting surface enhanced wireless network
  via joint active and passive beamforming,'' {\em IEEE Transactions on
  Wireless Communications}, vol.~18, no.~11, pp.~5394--5409, 2019.

\bibitem{ning2020beamforming}
B.~Ning {\em et~al.}, ``Beamforming optimization for intelligent reflecting
  surface assisted mimo: A sum-path-gain maximization approach,'' {\em IEEE
  Wireless Communications Letters}, vol.~9, no.~7, pp.~1105--1109, 2020.

\bibitem{xiong2023ris}
R.~Xiong {\em et~al.}, ``{RIS}-aided wireless communication in real-world:
  Antennas design, prototyping, beam reshape and field trials,'' {\em arXiv
  preprint arXiv:2303.03287}, 2023.

\bibitem{stutzman2012antenna}
W.~L. Stutzman and G.~A. Thiele, {\em Antenna theory and design}.
\newblock John Wiley \& Sons, 2012.

\end{thebibliography}

\section*{Biographies}
\vspace{-40pt}
\begin{IEEEbiographynophoto}{Rujing Xiong}
[S'22] (rujing@hust.edu.cn) received his B.S. degree in bioinformatics from Zhengzhou University, China in 2017, and his M.S. degree in electronics and communication engineering from Central South University in 2020. Currently, he is pursuing a Ph.D. degree at Huazhong University of Science and Technology. 
\end{IEEEbiographynophoto}
\vspace{-50pt}
\begin{IEEEbiographynophoto}{Jianan Zhang}
[S'22] (zhangjn@hust.edu.cn) received her M.S. degree in Information and Communication Engineering from Shanxi University, Taiyuan, China, in 2021. She is currently pursuing a Ph.D. degree with Huazhong University of Science and Technology. 
\end{IEEEbiographynophoto}
\vspace{-50pt}

\begin{IEEEbiographynophoto}{Xuehui Dong}
[S'22] (dong\_xh@hust.edu.cn) received his B.S. degree in Communication Engineering at Huazhong University of Science and Technology, China, in 2021, where he is currently pursuing a Ph.D. degree with the School of Electronic Information and Communications.
\end{IEEEbiographynophoto}
\vspace{-50pt}
\begin{IEEEbiographynophoto}{Zhengyu Wang}
[S'22] (wangzhengyu@hust.edu.cn) received the B.S. degree in communication engineering from Central South University, Changsha, China, in 2020.  She is currently pursuing a Ph.D. degree at Huazhong University of Science and Technology, China.
\end{IEEEbiographynophoto}
\vspace{-50pt}
\begin{IEEEbiographynophoto}{Junshuo Liu}
[S'23]  (junshuo\_liu@hust.edu.cn) received a B.S. degree from Nanjing University of Posts and Telecommunications in 2018, and an MS degree from King's College London, U.K., in 2019. He is currently pursuing a Ph.D. degree at Huazhong University of Science and Technology, China. 
\end{IEEEbiographynophoto}

\vspace{-50pt}
\begin{IEEEbiographynophoto}{Wei Yang}
[S'23] (yangwei\_eic@hust.edu.cn) received his B.E. degree from Wuhan University of Science and Technology, in 2021. He is currently enrolled at Huazhong University of Science and Technology for his M.S.
\end{IEEEbiographynophoto}
\vspace{-50pt}
\begin{IEEEbiographynophoto}{Tiebin Mi}
[M] (mitiebin@hust.edu.cn) received the B.E. degree in computer science from Xidian University, China, in 2002, and the Ph.D. degree in electrical engineering from Institute of Acoustics, Chinese Academy of Sciences, China, in 2010. He is currently a Lecture (Assistant Professor) with  Huazhong University of Science and Technology, China. 
\end{IEEEbiographynophoto}
\vspace{-50pt}
\begin{IEEEbiographynophoto}{Wenbo Huang}
[S'21] (eric\_huang@hust.edu.cn)received a B.E. degree in Electronic Engineering from Huazhong University of Science and Technology, China, in 2021, where he is currently pursuing a Ph.D. degree. 
\end{IEEEbiographynophoto}
\vspace{-50pt}
\begin{IEEEbiographynophoto}{Robert Caiming Qiu}[F] (caiming@hust.edu.cn) received a Ph.D. degree in electrical engineering from New York University. He joined the School of Electronic Information and Communications, Huazhong University of Science and Technology, China, as a full Professor since 2020. 
\end{IEEEbiographynophoto}

\end{document}